\documentclass{aa}
\usepackage{graphicx}
\usepackage{txfonts}
\begin{document}
   \title{Numerical simulations of coronal magnetic field loop evolution}

   \author{T.G. Yelenina,
          \inst{1}
          G.V. Ustyugova\inst{1}
          \and
          A.V. Koldoba\inst{2}
          }

   \offprints{T.G. Yelenina \email{elen@keldysh.ru}}

   \institute{Keldysh Institute for Applied Mathematics (KIAM), Russian Academy of
Sciences, Miusskaya square 4, 125047 Moscow, Russia \\
              \email{elen@keldysh.ru}
             ; \email{ustyugg@rambler.ru}
        \and
             Institute of Mathematical Modelling (IMM), Russian Academy of
Sciences, Miusskaya square 4a, 125047 Moscow, Russia\\
             \email{koldoba@spp.keldysh.ru}
             }

   \date{will be inserted by hand later }

  \abstract
   {}
   {We present the results of the numerical simulations of the
interaction between a magnetized star and an imperfectly
conducting accretion disk.}
   {To analyze the ''star--disk'' interaction we numerically investigate the MHD equations used Godunov-type
   high resolution numerical method.}
   {It was found that the ''star -- disk'' interaction occurs with quasi-periodic
reconnection of the magnetic field coronal loops and plasmoid
ejections. In the case of the perfect disk conductivity the
evolution of the coronal magnetic field leads to the periodic
outflow of angular momentum from the disk. In the case of an
imperfectly conducting disk the configuration of the magnetic
field is formed such that the disk angular momentum carried by
magnetic field gets balanced by angular momentum carried by
matter.}
   {}


   \maketitle
%

\section{Introduction}

This paper studies the evolution of the coronal magnetic field
linked with a magnetized star and its accretion disk. We suggest
that the plasma differential rotation along magnetic field lines
is the reason of a ''star--corona--disk'' system evolution. We
suppose that the magnetic field lines are frozen in the perfectly
conducting coronal plasma. The differential rotation leads to the
generation of the magnetic field toroidal component. Magnetic
pressure increases in corona inner part and plasma is pushed
towards the outer part together with the magnetic field lines. As
a result there is deformation or even opening of poloidal
magnetic field lines adopting a new configuration. The type of
this new configuration is determined by several factors. One of
them is the electrical conductivity of the relatively cold disk
plasma.

We assume in this model that the imperfect plasma conductivity is
essential only in the disk and is determined by velocity
turbulent fluctuations. The value of the turbulent magnetic
diffusivity we consider as a free parameter of the problem. To
obtain an acceptable range for this parameter, it is supposed
that the coefficient of the turbulent magnetic diffusivity agrees
in order of magnitude with the turbulent viscosity accepted in
the standard $\alpha$-model of Shakura--Sunyaev accretion disk
(Shakura \& Sunyaev \cite{Shakura}).

We consider that the disk is formed by relatively dense and cold
matter. The disk is Keplerian and sound speed is much less than
the Keplerian one. It means that disk is geometrically thin. In
this model, the disk is considered as an infinitely thin,
conductive plane. It should be mentioned that disk has a
complicated structure and its interaction with the magnetic field
does not reduce to magnetic compression and magnetic field lines
slippage relatively to matter (Balbus et al. \cite{Balbus}).

Besides the dynamics of the magnetic field, it is worthy to know
its configuration after the opening of the magnetic field lines.
This configuration defines the disk and the magnetic field
evolution at large time scales. The main factor influencing this
evolution is the rate of angular momentum transfer from the disk.
The important role in this process belongs also to the magnetic
field (Ferreira \cite{Ferreira1}).

A sufficiently strong magnetic field can lead also to the
formation of a matter outflow from the disk to the corona. In case
of a thin Keplerian disk, a criterion for the ''wind'' formation
beginning has been given by Blandford \& Payne (\cite{Payne}). To
generate this outflow from a Keplerian disk, the magnetic field
line should be inclined to the rotation axis with an angle of more
than $30^{\circ}$. In that case a magnetic field line plays the
role of a ''rail'' along which matter leaves the disk. Thus the
final magnetic field configuration determines strongly both the
rate of disk accretion, through the rate of angular momentum
outflow, and the rate of matter outflow along the magnetic field
lines inclined to the rotation axis.

There are lots of papers concerning the evolution of the magnetic
field in interaction with a disk. We would like to mention some of
them.  Hayashi et al. (\cite{Hayashi}) presented resistive MHD
simulations with diffusive accretion disk and dipole stellar
magnetic field topology. Kuwabara et al. (\cite{Kuwabara})
introduced resistivity to simulate the effects of turbulent
magnetic diffusivity. Romanova et al. (\cite{Romanova2}) studied
the disk accretion to a rotating magnetized star with an aligned
dipole moment and associated funnel flows, they described the
''star--disk'' interaction for the cases of fast and slowly
rotating star. These simulations included a treatment of the disk
vertical structure. Some papers presented the results of ideal MHD
simulations considered the disk like a boundary condition.
Ustyugova et al. (\cite{Ustyugova2}) for initially split-monopole
magnetic field configuration, Ouyed \& Pudritz (\cite{Pudritz})
for dipole topology, Krasnopolsky et al. (\cite{Krasnopol})
studied MHD outflows from the disk. Fendt \& Elstner
(\cite{Fendt1}, \cite{Fendt2}) studied the ''star--disk''
interaction, they observed the opening of magnetic field lines and
outflows. Fendt \& \v{C}emelji\'{c} (\cite{Fendt3}) studied the
jet formation and propagation by used the resistive MHD equations.

Shu et al. (\cite{Shu}) presented steady state dynamics of
accretion from viscous and imperfectly conducting disk. Ferreira
(\cite{Ferreira1}) gave self-similar solution for the resistive
disk for stationary MHD equations. Uzdensky (\cite{Uzdensky}) gave
a review of modern theoretical scenario of the star--disk system
interaction, including both stationary and time-dependent ones.

In this paper we consider the evolution of the magnetic field
configuration from an initial dipole-like topology into the final
one. We analyze the time-dependent ideal MHD equations using the
high resolution Godunov-type method. We consider for a setup
model of ''star--disk'' system that the generated flow is
axisymmetric and symmetric in relation with the equatorial plane.
The dependence of the disk angular momentum outflow rate on the
disk surface electrical conductivity is investigated.

The structure of the paper is as follows. After the introduction
we propose the model and the evaluation of the disk surface
magnetic diffusivity. Then we present the numerical method and
describe our results for number disk surface magnetic
diffusivity. Finally we formulate the main conclusions.


\section{Statement of the problem}

   \begin{figure}
   \centering
   \includegraphics[width=7cm]{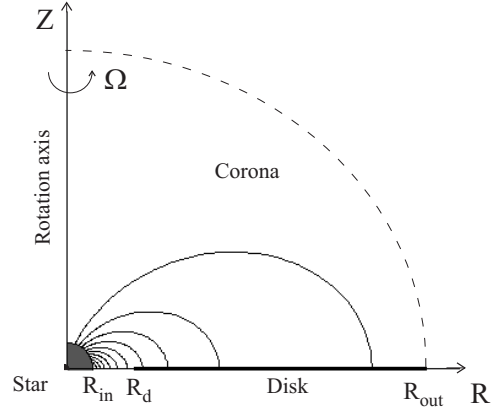}
      \caption{Sketch of a magnetically linked ''star--disk''
      system. The solid lines show the magnetic field lines.
      $R_{\mathrm{in}}$, $R_{\mathrm{out}}$ are the edges of the computational domain.
      The star is inside the computational domain, $R_*<R_{\mathrm{in}}$. }
         \label{FigDip}
   \end{figure}

Let us consider a ''star--corona--disk'' system governed by the
magnetic field (see Fig.~\ref{FigDip}). In our model the star has
mass $M_*$, magnetic moment $\mu_*$ and angular velocity
$\Omega_*$. It is assumed that the star rotation axis aligns with
its magnetic axis $z$. The disk rotating around the star is
considered to be infinitely thin and located at $z=0$. It is
supposed that the disk is rotating following Keplerian orbits
around the star and the disk is imperfectly conducting. The
particle velocity in a Keplerian orbit with radius $r$ is
$V_{\mathrm{k}}=\sqrt{GM_*/r}$, where
$G=0.667\cdot10^{-7}\mbox{cm}^3\mbox{g}^{-1}\mbox{sec}^{-2}$ is
gravitational constant. Thus, the disk is differentially rotating.

The coronal plasma electro-conductivity is big enough, so we can
describe the flow by the system of the ideal MHD equations
\begin{equation}
 \label{eq3-1}
    \begin{array}{l}
    \displaystyle{
    \frac{\partial \rho}{\partial \, t} + \nabla \cdot (\rho \, {\vec u}) = 0,}\\
 \\ \displaystyle{ \frac{\partial \rho \, {\vec u}}{\partial \, t} + \nabla \, {\tens T} = \rho \, {\vec g},}  \\
 \\  \displaystyle{ \frac{\partial {\vec B}}{\partial \, t} - \nabla
   \times \, ({\vec u} \times {\vec B}) = 0, }  \\
 \\  \displaystyle{ \frac{\partial \rho \, S}{\partial \, t} +
    \nabla \cdot (\rho \, S \, {\vec u}) = 0, }  \\
 \\   \nabla\,{\vec B} = 0.
    \end{array}
\end{equation}
Here $\displaystyle{{\tens T}_{ik} = \rho\,u_i u_k +
p\,\delta_{ik} + \frac{1}{4 \pi}\left(-B_i B_k +
\frac{B^2}{2}\delta_{ik}\right)}$ is the stress tensor; $\vec u$
is the plasma velocity; $\vec B$ is the magnetic field; $\rho$ and
$p$ are the plasma density and pressure; $S=p/\rho^{\gamma}$ is
the entropy function; $\gamma$ is the adiabatic index; $\vec
g=-\nabla\Phi_{\mathrm{g}}$ is the gravitational acceleration;
$\Phi_{\mathrm{g}}=-G M_*/R $ is the star gravity potential; $R$
is the distance from the gravitating center.

The system (\ref{eq3-1}) is solved in spherical coordinates
$(R,\varphi,\theta)$, with $\theta$ being the polar angle with the
symmetry axis. Velocity $\vec u$ and magnetic field $\vec B$ have
all their components $\vec u=(u,v,w)$ and $\vec
B=(B_R,B_{\varphi},B_{\theta})$.

The aim of this paper is to investigate the character of the
magnetic field evolution and field topology depending on the disk
surface magnetic diffusivity $\zeta=c^2/2\pi\lambda$ ($\lambda$ is
surface electric conductivity, $c$ is the speed of light) at
large time scales. As it will be shown below, it used in boundary
conditions set on the equatorial plane $z=0$.

\subsection{Dimensionless variables and typical quantities for T
Tauri stars}

The dimensionless form of (\ref{eq3-1}) is received in standard
way. As distance scale, $R_0$, we take one third of the distance
from the star center to the inner edge of the disk. Thus, in
dimensionless units the disk inner edge is
$R_{\mathrm{d}}=3R_{\mathrm{in}}$. The inner radius of the
computational domain in dimensionless units is
$R_{\mathrm{in}}=1$. The time and the velocity scales are chosen
that in dimensionless units $GM_*=1$. This requirement yields
$t_0=\sqrt{R_0^3/GM_*}$ as the time scale and $v_0=R_0/t_0$ as the
velocity scale. As a magnetic field scale, $B_0$ is taken and then
density, pressure and magnetic moment scales are
$\rho_0=B_0^2/v_0^2$, $p_0=B_0^2$, $\mu_{*0}=B_0 R_0^3$.

As typical quantities of T Tauri stars we adopt standard values
like a star mass $M_*=0.8 M_{\sun}=1.6 \cdot 10^{33}$ g, and
$R_d=9R_{\sun}=5.4 \cdot 10^{11}$ cm. Therefore the distance scale
is $R_0 = R_{\mathrm{d}}/3 = 1.8 \cdot 10^{11}$ cm, the time and
velocity scales are $0.74 \cdot 10^{4}$ s and $2.43 \cdot 10^{7}$
cm s$^{-1}$ respectively. The Keplerian rotation period at the
disk inner edge is $8.3$ days. The simulation region size is
$1.134 \cdot 10^{12}$ cm.

The star magnetic moment is taken such that the magnetic field on
the star surface is 300 G, so $B_0=6.5$ G. Thus, on the disk inner
part, the dipole magnetic field is 2.4 G, $R_*=3/5
R_{\mathrm{in}}$, $\mu_*=10$ and the magnetic moment is $3.8 \cdot
10^{35}\mbox{G}\,\mbox{cm}^3$. The density scale is $1.44 \cdot
10^{-14}\mbox{g} \, \mbox{cm}^{-3}$, typical for the disks around
T Tauri stars.

\subsection{Evaluation of the electric conductivity in the disk}

We suggest that turbulent diffusion of magnetic field is
determined by the same processes that determine turbulent
viscosity, which leads to angular momentum transport in the disk.
Thus, it is assumed that turbulent magnetic diffusivity
$\eta_{\mathrm{t}}$ is in the order of turbulent viscosity like in
the Shakura--Sunyaev model (Shakura \& Sunyaev 1973):
$\eta_{\mathrm{t}} = \alpha_{\mathrm{t}} \, c_{\mathrm{s}} \, h$,
where $c_{\mathrm{s}}$ is the sound speed in the disk, $h$ is the
disk half-thickness, $\alpha_{\mathrm{t}}$ is the dimensionless
coefficient varying, according to Balbus (\cite{Balbus2003}), in
this range $0.01 \div 0.6$.

Under hydrostatic equilibrium, the Keplerian disk half-thickness
$h$ can be found from the relation:
$(h/r)^2+b(h/r)-(c_{\mathrm{s}}/V_{\mathrm{k}})^2=0$, where $b
\equiv r(B_r^2+B_{\varphi}^2)/(4\pi\Sigma V_{\mathrm{k}}^2)$,
$\Sigma$ is the surface density, $V_{\mathrm{k}}$ is Keplerian
velocity, $B_r$, $B_{\varphi}$ are the magnetic field components
(Bisnovatyi--Kogan \& Lovelace \cite{Lovelace1}). In any case,
even without taking into account the magnetic compression, the
disk half-thickness satisfies the inequality $h \la
c_{\mathrm{s}}/\Omega_k$, where $\Omega_{\mathrm{k}}$ is Keplerian
angular velocity. Therefore, the turbulent magnetic diffusivity
becomes $\eta_{\mathrm{t}} \la \alpha_{\mathrm{t}}
c_{\mathrm{s}}^2/\Omega_{\mathrm{k}}$.

The sound speed is much less than the Keplerian one is due to the
disk is cold. It means that $h \la
c_{\mathrm{s}}/\Omega_{\mathrm{k}} \ll r$, i.e. disk is
geometrically thin. In this model, the disk is considered as an
infinitely thin, conductive plane.

The turbulent electro-conductivity $\sigma_{\mathrm{t}} = c^2/(4
\pi \eta_{\mathrm{t}}) = c^2/(4 \pi \alpha_{\mathrm{t}}
c_{\mathrm{s}} h)$ is associated with the magnetic diffusivity
$\eta_{\mathrm{t}}$. The surface disk conductivity is $\lambda =
\int \sigma_{\mathrm{t}} d z \sim 2 h \sigma_{\mathrm{t}} = c^2/(2
\pi \alpha_{\mathrm{t}} c_{\mathrm{s}} )$, and the surface
magnetic diffusivity is
\[
\zeta = \frac{c^2}{2 \pi \lambda} = \alpha_{\mathrm{t}}
c_{\mathrm{s}} = \alpha_{\mathrm{t}}
\left(\frac{c_{\mathrm{s}}}{V_{\mathrm{k}}}\right)V_{\mathrm{k}} .
\]

In thin accretion disks $c_{\mathrm{s}}/V_{\mathrm{k}} = h/r\ll
1$. Thus, an acceptable coefficient of the magnetic surface
diffusivity is $\zeta = (0.01 \div
0.6)(c_{\mathrm{s}}/V_{\mathrm{k}})V_{\mathrm{k}}$.

\subsection{Initial conditions}

We suggest that, at the initial time, the stellar magnetic field
with dipole-like topology and magnetic moment $\mu_*$ penetrates
the corona and the disk. So, the components of the magnetic field
${\vec B}$ are
\[
B_R = \frac{2 \mu_* \cos \theta}{R^3}, \, B_{\theta} = \frac{\mu_*
\sin \theta}{R^3},\, B_{\varphi}=0.
\]
We suppose that at the initial time $t=0$, the matter of the
corona and the disk is in mechanical equilibrium with the
force-free dipole magnetic field, i.e. gravitational force is
balanced with ''centrifugal'' force (liquid particle
acceleration) and pressure gradient.

The momentum equation for the system (\ref{eq3-1}) taking into
account that particles follow circular orbits is
\begin{equation}
 \label{eq3-3}
    -\omega^2 r {\vec e}_r+\frac{1}{\rho}\nabla p=-\nabla\Phi_g ,
\end{equation}
where ${\vec e}_{\mathrm{r}}$ is a unit vector with the direction
of the cylindrical radius $r=R\sin\theta$, that on disk surface it
is $\theta=\pi/2$ and $r=R$.

This is due to the magnetic pressure ${\vec B}^2/8\pi$ of the
dipole-like field ${\vec B}$ decreases like $1/R^6$. Also, the
most of the simulation region is occupied by a relatively dense
plasma, where gas pressure dominates. This dense plasma prevents
opening of the magnetic field lines. Let us consider that density
$\rho$ is function not only of pressure $p$ but also of
cylindrical radius $r$: $\rho=\rho(r,p)$. Let us denote
$V(p,r)=1/\rho$. The momentum equation along the $z$ axis (the
projection of equation (\ref{eq3-3}) to the $z$ axis) is
\[
V(p,r)\frac{\partial p}{\partial z}+\frac{\partial
\Phi_g}{\partial z}=0.
\]
Integrating it by $z$ and suggested that $p\to 0$ under
$z\to\infty$ and $\Phi_g \to 0$) we finally get that
\begin{equation}
 \label{eq3-5}
    \int \limits_0^p V(p',r) d p' + \Phi_g = 0.
\end{equation}
The integral is assumed to converge on its interior limit.

Let us consider now the momentum equation (\ref{eq3-3}) along the
radial direction
\begin{equation}
 \label{eq3-6}
    -\omega^2 r + V(p, r) \frac{\partial p}{\partial r}
    +\frac{\partial \Phi_{\mathrm{g}}}{\partial r} = 0,
\end{equation}
After differentiating equation (\ref{eq3-5}) by $r$ and
subtracting (\ref{eq3-6}) from it we get
\begin{equation}
 \label{eq3-8}
    \omega^2 r + \int \limits_0^p
    \frac{\partial V }{\partial r} d p' = 0.
\end{equation}
Thus, given the function $V(p,r)$, the relation (\ref{eq3-5})
defines the function $p\,(r,z)$ and relation (\ref{eq3-8}) --
function $\omega(r,z)$.

If $V(p,r)=k(r)/p^{\alpha}$, $\alpha = {\rm const} < 1$ (this
condition is essential for the convergence of the integral in left
part of equation (\ref{eq3-5}) as $p\to 0$), then equations
(\ref{eq3-5}), (\ref{eq3-8}) take the form
\[
\left\{ {
\begin{array}{l}
\displaystyle{
    \frac{k p^{\,1 - \alpha}}{1 - \alpha} + \Phi_g = 0, } \\
\displaystyle{
    \frac{p^{\,1-\alpha}}{1 - \alpha} k' + \omega^2 r = 0.  }
  \end{array}
} \right.
\]
Integrating it, we found the function $k(r)$
\begin{equation}
 \label{eq3-81}
\ln k = -\frac{1}{G M_*} \int \Omega^2(r) r^2 d r .
\end{equation}
Here $\Omega(r)=\omega(r,0)$ is the angular velocity at equatorial
plane $z=0$.

The angular velocity at the equatorial plane was chosen in the
following way ($r_1$, $r_2$ are the parameters of the problem and
$r_1=2 R_{\mathrm{in}}$, $r_2=3 R_{\mathrm{in}}$):
\begin{equation}
 \label{eq3-82}
\Omega(r)=\left\{{
\begin{array}{l}
  \displaystyle{  \Omega_{\ast} = \sqrt{GM_*/r_1^3}, \,\,\,  0 < r < r_1, }  \\
 \displaystyle{   \Omega_{\ast}+\frac{\sqrt{GM_*/r_2^3}-\Omega_{\ast}}{r_2-r_1 }(r-r_1), \,\,\, r_1 <
    r < r_2,} \\
  \displaystyle{  \sqrt{GM_*/r^3}, \,\,\, r > r_2. } \\
\end{array} } \right.
\end{equation}

As a result, the distributions of the pressure $p(r,z)$, density
$\rho(r,z)$ and angular velocity $\omega(r,z)$ are obtained
\[
\begin{array}{l}
\displaystyle{p(r,z) = \left( \frac{1 - \alpha}{k(r)}\frac{G
M_*}{\sqrt{r^2 + z^2}} \right)^{1/(1-\alpha)}, \, \rho(r,z) =
\frac{p^{\,\alpha}(r,z)}{k(r)}, }\\ \displaystyle{ \omega =
\sqrt{\frac{p^{\,1 - \alpha}}{\alpha - 1}\frac{k'(r)}{r}} .}
\end{array}
\]

\subsection{Boundary conditions}

The surface currents in the disk (at $z = 0$) lead to a
discontinuity in the magnetic field disk-tangential components.
Mathematically it means that the following condition is fulfilled
(Landau \& Lifshitz, 1982)
\begin{equation}
 \label{eq3-11}
    {\vec n}\times({\vec B}^+ -{\vec B}^-)=\frac{4\pi}{c}{\vec i},
\end{equation}
where ${\vec B}^+$, ${\vec B}^-$ are the magnetic field under and
over the disk respectively, ${\vec n}$ is the unit normal to the
disk and directed downward in cylindrical coordinates. Since we
suggest that MHD-flow is symmetric in relation to equatorial
plane and consider the problem in the upper half-space, then
${\vec B}^+=-{\vec B}^- = {\vec B}$ and (\ref{eq3-11}) gives the
following
\begin{equation}
 \label{eq3-111}
{\vec n}\times{\vec B} + \frac {2 \pi}{c} {\vec i} = 0.
\end{equation}
Substituting expression for the surface current in the disk ${\vec
i}$ which is due to the Ohm's law at the comoving frame $ {\vec i}
= - \lambda ({\vec u}- {\vec V})\times {\vec B} / c $ into the
equation (\ref{eq3-111}) for current, finally we get
\begin{equation}
 \label{eq3-112}
({\vec u} - {\vec V} - \zeta {\vec n})\times{\vec B} = 0.
\end{equation}
Assuming that the angular velocity $\omega$ and $r, \,
\varphi$-components of electric field in comoving frame change
negligibly in $z$-direction on scales of the order of the disk
thickness $2h$, we obtain for the disk surface electric
conductivity $\lambda = \int \sigma_{\mathrm{t}} d z \sim 2 h
\sigma_{\mathrm{t}}$.

In spherical coordinates the tangential components of
(\ref{eq3-112}) are
\begin{equation}
 \label{eq3-114}
\begin{array}{l}
    (v - V_{\mathrm{k}})B_{\theta} - (w - \zeta)B_{\varphi} = 0, \\
    u B_{\theta} - (w - \zeta) B_R = 0 .
\end{array}
\end{equation}
The relations (\ref{eq3-114}), (expressing Ohm's law for surface
current in the disk), give two boundary conditions in the
equatorial plane under $\theta=\pi/2$ ($z=0$).

The real disk has some vertical (in $z$-direction) structure that
is not considered here. It is essential that thermodynamic
parameters of the disk plasma change in the vertical direction
turning smoothly to the corona ones. So, the boundary conditions
can be set arbitrarily, based on some physically reasonable
assumptions.

We assume that matter flows from the disk to the corona at small
velocity. It could be expected that plasma leaves the disk with
the velocity less than the slow magnetosonic one. Let us accept
that the $z$-component of the velocity is a fraction $\alpha_c$ of
the cusp one in this direction. The cusp velocity is not greater
than slow magnetosonic one and has the same direction. The
condition (\ref{eq3-115}) guarantees that the plasma outflow does
not exceed the slow magnetosonic one, giving the third boundary
condition for $\theta=\pi/2$ (${\vec a}={\vec B}/\sqrt{4\pi\rho}$
is the Alfv\'{e}n velocity, parameter $\alpha_c(R)<1$):
\begin{equation}
 \label{eq3-115}
w+\alpha_c(R)\frac{c_{\mathrm{s}}|a_{\theta}|}{\sqrt{c_{\mathrm{s}}^2+a^2}}=0.
\end{equation}
The condition (\ref{eq3-115}) guarantees also that from any point
of the disk five characteristics come out. So, one should set
another two boundary conditions.

We assume that matter outflow from the disk does not change its
interior structure, i.e. the disk has sufficiently large mass,
energy and angular momentum. We assume that the specific entropy
of the matter flowing out from the disk does not vary with time
\[
S \big|_{\theta=\pi/2}=S_d(R).
\]

To get the equation describing the evolution of the magnetic flux
function on the disk (at $\theta=\pi/2$) let us use the induction
equation
\begin{equation}
 \label{eqind}
   \frac{\partial B_{\theta}}{\partial t}+\frac{c}{R}\frac{\partial R E_{\varphi}}{\partial R}=0.
\end{equation}
Taking into account that on the disk $B_{\theta}= {\partial
\Psi}/(R\,{\partial R})$ and Ohm's law is fulfilled, the azimuth
component of the surface current,
$E_{\varphi}+c\,B_R/(2\pi\sigma_t)=0$. Integrating (\ref{eqind}),
we found
\[ \frac{\partial
\Psi}{\partial t}+\zeta R B_R=0.
\]

Let us formulate the boundary conditions on the inner bound of the
simulation region under $R=R_{\mathrm{in}}$. As for boundary
conditions on equatorial plane, they are set for reasons of
physical rationality. The main factor to take into account is on
the one hand to choose arbitrary the position of the inner bound,
i.e. the value of the inner radius of the simulation region
$R_{\mathrm{in}}$, and, on the other hand, quickly (like $R^{-6}$)
increasing magnetic pressure of stellar dipole-like field.
$R_{\mathrm{in}}$ is chosen to be such that in some neighborhood
of the inner bound, the star magnetic field has dominant influence
on the plasma dynamics. In other words, in this region Alfv\'{e}n
velocity is much more than both the gas sound speed and the
Keplerian one. On the other hand, the choice of a very small
$R_{\mathrm{in}}$ is not reasonable from the computational point
of view, because, into the simulation region, in this case
includes parts of the magnetosphere where Alfv\'{e}n velocity is
too large. It leads to an essential decreasing of time integration
step. Since in the considered model it is assumed that magnetic
field lines are frozen into the star surface rotating with angular
velocity $\Omega_*$, on the inner bound of the simulation region
(and under it, right on the star surface) plasma moves along the
rotating magnetic field lines. In the frame of the rotating star
(on inner bound) plasma velocity vector is parallel to the
magnetic field one. Since transforming to the rotating frame
${\vec B }$ does not change, but ${\vec u}$ transforms into ${\vec
u}-V_*{\vec e}_{\varphi}$, where
$V_*=\Omega_*R_{\mathrm{in}}\sin\theta$. This boundary condition
can be written in the following way ($R=R_{\mathrm{in}}$):
\begin{equation}
 \label{eq3-116}
({\vec u}-V_*{\vec e}_{\varphi}) \times {\vec B}=0.
\end{equation}
Condition (\ref{eq3-116}) implies also that in the rotating frame,
the electric field in the inner bound is zero. The conditions
(\ref{eq3-116}) give two boundary conditions under
$R=R_{\mathrm{in}}$.

On the outer bound, at $R=R_{\mathrm{out}}$, ''free'' boundary
conditions are set. Such conditions should not influence on the
solutions inside simulation region preventing poloidal magnetic
field lines opening.

On the rotation axis, although it is not a bound, symmetry of the
flow conditions for this axis are set:
\[
v=w=0,\;B_{\varphi}=B_{\theta}=0,\;\Psi=0\;\mbox{ under }\;
\theta=0.
\]

\section{Numerical method and results}

For the numerical integration of the ideal MHD equations
(\ref{eq3-1}), we use Godunov-type conservative high resolution
scheme (Kulikovskii et al. \cite{Kul}, Yelenina \& Ustyugova
\cite{ElenUstyug2}). To guarantee divergence--free magnetic field
we apply the same procedure as T\'{o}th (\cite{Toth}).

The system of equations (\ref{eq3-1}) is integrated numerically in
the region $R_{\mathrm{in}} < R < R_{\mathrm{out}}$, $0 < \theta <
\pi/2$. We take a non-uniform grid in the radial direction, and
uniform in the polar angle: $N_{\theta} = 60$, $N_{R} = 60$. The
time step of the integration $\tau$ is restricted by Courant
condition.

To verify the method and results we also used two grids:
$120\times60$ and $240\times120$ and first order numerical method
to estimate the numerical diffusivity of the scheme. To satisfy
the goals of this paper, i.e. to investigate the regime of
interaction between the magnetized star and the disk, it is turned
out that the grid size presented in the paper is quite enough.

\begin{figure}
   \centering
   \includegraphics[width=8cm]{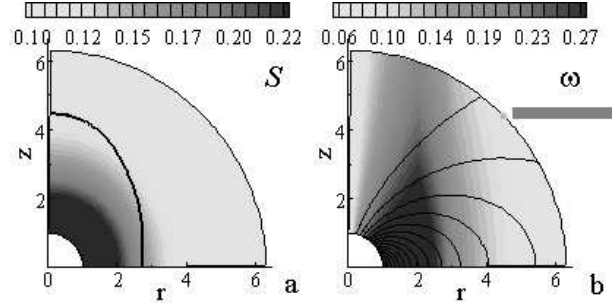}
      \caption{(a) Background colour shows the initial distribution of the specific entropy $S(r,z)$,
      solid line presents the magnetic parameter $\beta=1$.
      (b) Background colour shows the initial distribution of the angular velocity
      $\omega(r,z)$, thin lines present the magnetic flux function $\Psi(r,z)$.}
         \label{ini}
   \end{figure}

\begin{figure*}
   \centering
   \includegraphics[width=12cm]{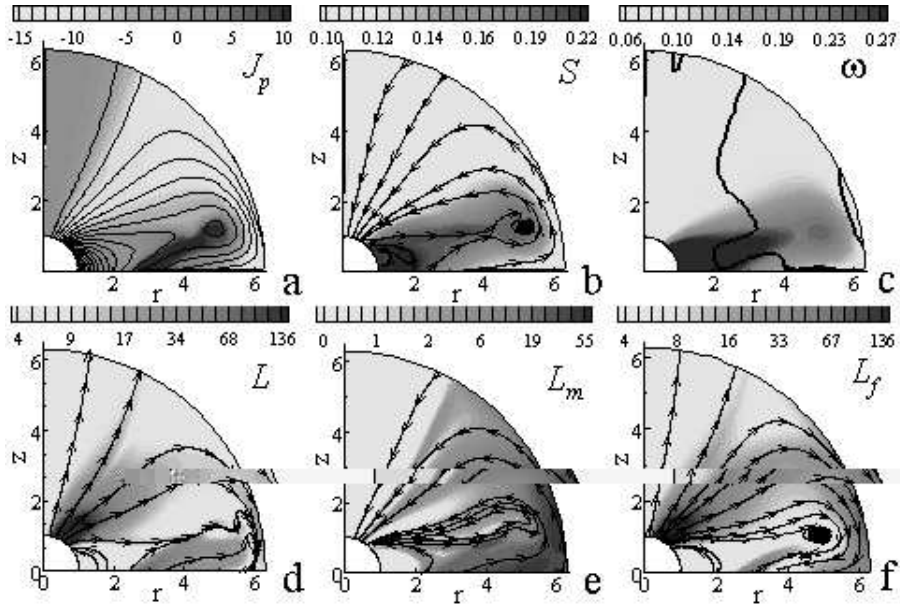}
      \caption{Flow distributions for $\zeta=0.0$ at $t=50$. (a) Background colour shows the poloidal current
      $J_p$. The magnetic field lines are shown by firm lines. There is a poloidal current in the corona, forming a
double current sheet (shown by background dark colour region). (b)
Background colour shows the specific entropy $S$. Streamlines
present the matter current lines. Plasmoid has hot matter
surrounding it from the corona and moves up to the outer boundary.
(c) Background colour shows the plasma angular velocity $\omega$,
firm line present the level of the plasma parameter $\beta=1$. In
the neighborhood of the star poloidal current lines approximately
coincide with the poloidal field ones. Fig. (d)--(f) show the
distribution of the angular moment transport in the system. (d)
Background colour shows the magnitude of the angular momentum
flux, and streamlines show the direction of the system angular
momentum transport. (e) Streamlines show the angular momentum flux
direction carried by matter. The background colour shows the
magnitude of this vector. (f) Streamlines present the angular
momentum flux direction, carried by the magnetic field, its
magnitude is shown by colour. There are two regions of intense
momentum transport. One of them is nearby the star pole where
transport takes place due to magnetic stresses (f). The second one
is above the disk, in the area of its differential rotation where
the transport is caused by matter flow (e). }
         \label{mu0520}
   \end{figure*}

\begin{figure*}
   \centering
   \includegraphics[width=12cm]{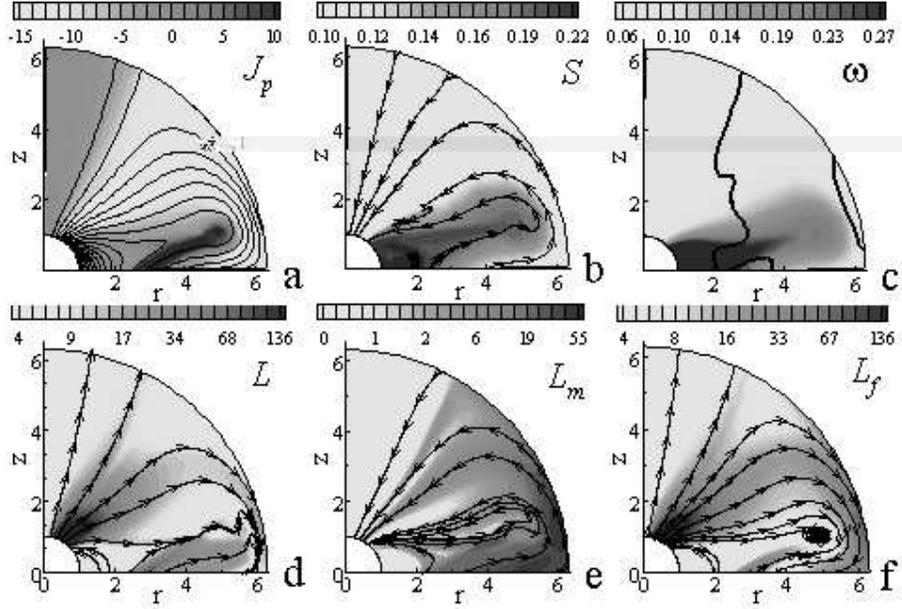}
      \caption{Flow distributions for $\zeta=0.001$ at $t=50$.
      (a) Background colour shows the poloidal current $J_p$. The magnetic field lines are shown by firm lines.
      (b) Background colour shows the specific entropy $S$. Streamlines present the matter current lines.
      (c) Background colour shows the plasma angular velocity $\omega$, firm line presents the plasma parameter level
$\beta=1$. Fig. (d)--(f) show the distribution of the angular
moment transport in the system. (d) Background colour shows the
magnitude of the angular momentum flux, and streamlines show the
direction of the system angular momentum transport. (e)
Streamlines show the angular momentum flux direction carried by
matter. The background colour shows the magnitude of this vector.
(f) Streamlines show the angular momentum flux direction, carried
by the magnetic field, its magnitude is presented by colour.  }
         \label{mu0503}
   \end{figure*}

\begin{figure*}
   \centering
   \includegraphics[width=12cm]{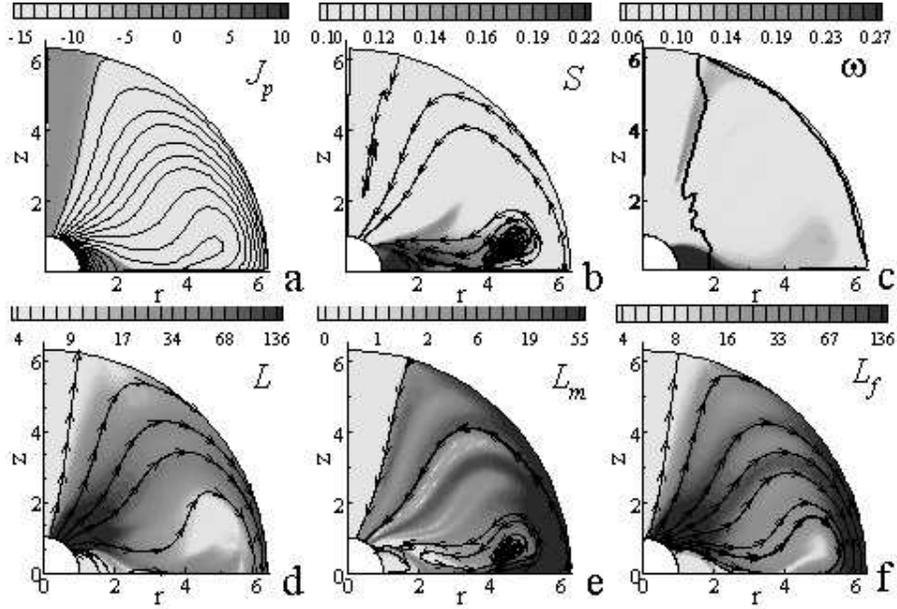}
      \caption{Flow distributions for $\zeta=0.005$ at $t=50$.
       (a) Background colour shows the poloidal current $J_p$. The magnetic field lines are shown by firm lines.
      (b) Background colour shows the specific entropy $S$. Streamlines present the matter current lines.
      (c) Background colour shows the plasma angular velocity $\omega$, firm line presents the plasma parameter level
$\beta=1$. Fig. (d)--(f) show the distribution of the angular
moment transport in the system. (d) Background colour shows the
magnitude of the angular momentum flux, and streamlines show the
direction of the system angular momentum transport. (e)
Streamlines show the angular momentum flux direction carried by
matter. The background colour shows the magnitude of this vector.
(f) Streamlines show the angular momentum flux direction, carried
by the magnetic field, its magnitude is presented by colour.   }
         \label{mu0505}
   \end{figure*}

\begin{figure*}
   \centering
   \includegraphics[width=12cm]{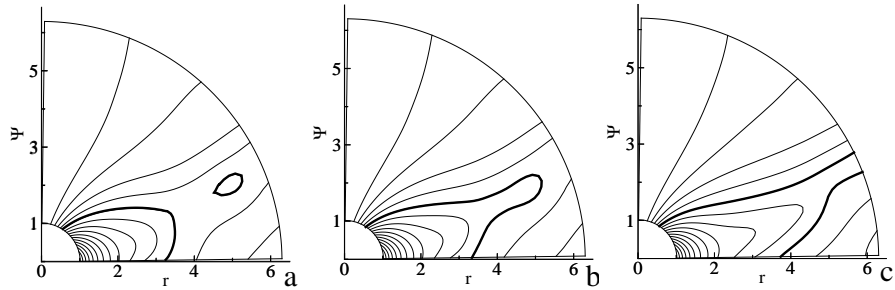}
      \caption{(a)--(c) Influence of the surface magnetic
diffusivity $\zeta$ on the magnetic field topology at $t=11$ when
the first reconnection took place for the case $\zeta=0$. Level
lines of the magnetic flux function $\Psi(r,z)$ are shown by thin
lines (magnetic field lines: on (a) for $\zeta=0.0$, on (b) for
$\zeta=0.001$, on (c) for $\zeta=0.005$, $\Psi=151.6$ is shown by
solid line. The more the magnetic diffusivity, the more the
distance from the axis and the disk where the plasmoid ejection
takes place. In case of imperfect disk conductivity, field lines
are no longer frozen into the disk and are shifted outwards. }
         \label{mu0506}
   \end{figure*}

\begin{figure*}
   \centering
   \includegraphics[width=12cm]{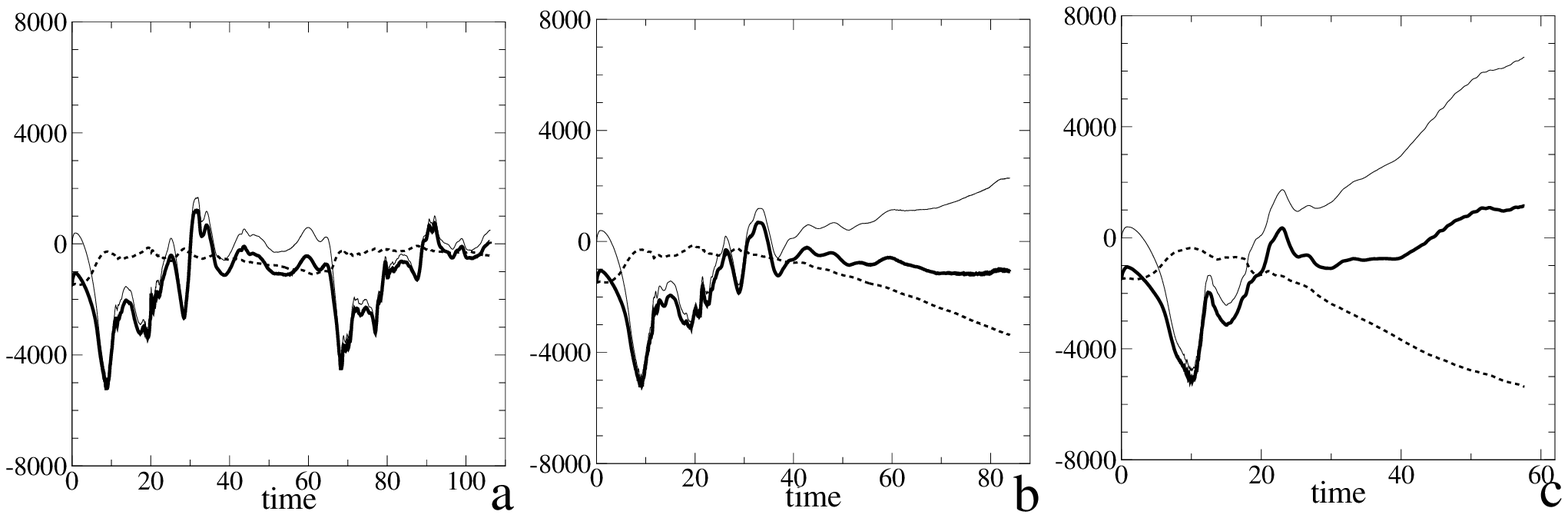}
      \caption{Time dependence of the angular momentum fluxes carried by
matter $L_m$, by magnetic field $L_f$ and total $L=L_m+L_f$ for
different $\zeta$: on (a) for $\zeta=0.0$, on (b) for
$\zeta=0.001$, on (c) for $\zeta=0.005$; $L_m$ is shown by dashed
line, $L_f$ is shown by thin line, $L$ is shown by solid line. For
$\zeta=0$ (a), the process of angular momentum transport from the
disk to the corona is quasi-periodic with period equal
approximately to ten rotation periods of the disk inner part.
Then, after the relaxation the process is resumed. For
$\zeta=0.001$ (b) and $\zeta=0.005$ (c), the magnetic lines
reconnection lasts until some time. The more the magnetic
diffusivity $\zeta$, the earlier the plasmoids ejection ends. The
reconnection of magnetic field lines takes place at the time
corresponding to the maximum angular momentum outflow carried by
magnetic field from the disk. In contrast to a case for
$\zeta=0$, the activity of the reconnection slows down and the
system relaxes to such a state that no new plasmoids are
generated. The disk angular momentum carried by magnetic field
gets balanced by angular momentum carried by matter, in
particular, $|L|\ll|L_f|$. }
         \label{mu0507}
   \end{figure*}

For mathematical simulation of the interaction between a
magnetized star and its accretion disk we performed several runs
for different surface magnetic diffusivity $\zeta$.
Fig.~\ref{ini}--\ref{mu0507} present the results for the following
values of parameter $\zeta$: 0, 0.001, 0.005.

Fig.~\ref{ini} shows the initial system configuration at time
moment, $t=0$. The background color in Fig.~\ref{ini}a shows the
plasma specific entropy distribution $S(r,z)$, thick line
corresponds to the plasma parameter $\beta=8\pi\,p/B^2=1$. The
background color on Fig.~\ref{ini}b shows the plasma angular
velocity distribution $\omega(r,z)$, streamlines show magnetic
field lines -- magnetic flux function $\Psi(r,z)$.

The evolution of the coronal magnetic field loops in the ''star --
disk'' system depends on the surface magnetic diffusivity $\zeta$.
One can pick out the characteristic features which are essential
for this process. For all the cases, poloidal field lines are
pulled out and reconnected periodically (approximately each tenth
rotation period of the disk inner bound). Then, a magnetic field
toroidal component is generated. After reconnection, a plasmoid is
formed. It is surrounded by closed poloidal magnetic field lines
along which the poloidal electric current is running. Plasmoid is
determined by a strong toroidal magnetic field and a low gas
pressure. The angular velocity inside the plasmoid is different
from the corona one. $t=50$ is chosen for presenting the
simulation results (the time is measured in disk inner bound
rotation period). Up to that moment, several reconnections of
magnetic field lines took place, and next ejected plasmoid moves
outwards to the outer bound. Previous series of reconnections
already led to the opening of the field lines close to the
rotation axis. We should note that the reconnection of the
magnetic field lines originates due to the numerical magnetic
diffusivity. Nevertheless, we suppose that the reconnection and
plasmoid ejection takes place as well for real magnetic
diffusivity.

Fig.~\ref{mu0520}a--f show the distributions of some variables for
a magnetic diffusivity $\zeta=0$. The background colour on
Fig.~\ref{mu0520}a shows the distribution of the poloidal current
$J_p=R\sin\theta B_{\varphi}$. The magnetic field lines are shown
also in this plot by firm lines. It is clear from these pictures
that the initial configuration has essentially changed and now it
is not dipole--like. In the disk differential rotation region
($r>2$), field lines come out of its surface with large slope
angle (more than $30^{\circ}$), fulfilling the conditions for
matter outflow from the disk. A toroidal magnetic field is
generated as a result of the differential rotation, implying that
there is a poloidal current in the corona, forming a double
current sheet (background dark colour area on Fig.~\ref{mu0520}a).
In the neighborhood of the star poloidal current lines
approximately coincide with the poloidal field ones.

The background colour on Fig.~\ref{mu0520}b shows the distribution
of the specific entropy $S$. Streamlines present the matter
current lines. It is clear that the matter flows into the plasmoid
area from the star and the disk inner part. Plasmoid has hot
matter surrounding it from the corona and moves up to the outer
boundary.

The background colour on Fig.~\ref{mu0520}c shows the plasma
angular velocity distribution, firm line -- plasma parameter level
$\beta=1$. It is evident from comparing Fig.~\ref{mu0520}a and
Fig.~\ref{mu0520}c that the angular velocity is practically
constant along the field lines, especially nearby the star. It
means that in this area the generation of the toroidal magnetic
field does not take place, i.e. there are no poloidal electric
currents. The magnetosphere, rotating with constant angular
velocity, gives angular momentum to the plasmoid, twisting the
matter inside it.

Fig.~\ref{mu0520}d -- \ref{mu0520}f show the distributions of
several quantities that describe the angular moment transport in
the system. The angular momentum conservation equation can be
found from the continuity equation and is
\[ \frac{\partial \rho
l}{\partial t}+{\rm div}{\vec L}=0,
\]
here $\rho l= \rho uR\sin\theta$ is the angular momentum density.
The poloidal components of the angular momentum flux density are
\[
{\vec L}=R \sin \theta \left( \rho v {\vec u}_p -
\frac{B_{\varphi}{\vec B}_p}{4\pi}\right),
\]
where ${\vec u}_p$, ${\vec B}_p$ are the plasma poloidal velocity
and the poloidal magnetic field. The first term in the right part
describes angular momentum transported by matter, the second one
does by magnetic field.

The background colour on Fig.~\ref{mu0520}d shows the magnitude of
the angular momentum flux, and streamlines show the direction of
the system angular momentum transport. There are two areas of
intense momentum transport. One of them is nearby the star pole
where transport takes place due to magnetic stresses (see
Fig.~\ref{mu0520}f). The second one is above the disk, in the area
of its differential rotation where the transport is caused by
matter flow. These processes are presented on Fig.~\ref{mu0520}e,f
in more detail. On Fig.~\ref{mu0520}e, streamlines show the
angular momentum flux direction carried by matter. The background
colour shows the magnitude of this vector. On Fig.~\ref{mu0520}f
streamlines -- angular momentum flux direction, carried by the
magnetic field, by colour -- its magnitude. It is clear from the
Fig.~\ref{mu0520}f that angular momentum from the star is
transported by magnetic field mainly. Also there is intense
momentum transport inside the plasmoid.

The distributions for the same variables at $t=50$ are shown on
Fig.~\ref{mu0503}--\ref{mu0505} in case of a finite surface
magnetic diffusivity. Fig.~\ref{mu0503}a--\ref{mu0503}f correspond
to $\zeta=0.001$, Fig.~\ref{mu0505}a--\ref{mu0505}f --
$\zeta=0.005$.

The Fig.~\ref{mu0506}a--c show the influence of the magnetic
diffusivity on the magnetic field topology. Level lines of the
magnetic flux function $\Psi(r,z)$ for all cases (a: $\zeta=0$, b:
$\zeta=0.001$, c: $\zeta=0.005$) are chosen at $t=11$. At this
time, the first reconnection took place for the case $\zeta=0$. It
is clear (see Fig.\ref{mu0506}a--c) that the less the disk
conductivity (i.e. the more magnetic diffusivity), the more the
distance from the axis and the disk where the plasmoid ejection
takes place. In case of imperfect disk conductivity, field lines
are no longer frozen into the disk and, twisted by the star, they
are shifted outwards. We can say that the more the surface
magnetic viscosity $\zeta$, the slower the evolution takes place,
and the more seldom plasmoids are formed in corona.

The matter flowing out from the disk and the magnetic field both
take away the angular momentum from it. The whole angular momentum
flux taken away from the disk per unit time is
\[
L=L_m+L_f=-\int \rho ({\vec r}\times{\vec u}){\vec u}d {\vec S} +
\frac{1}{4\pi} \int ({\vec r}\times{\vec B}){\vec B} d {\vec S}.
\]
Here, the integration is over the disk surface, $d {\vec S}$ is
the element of disk surface directed outwards (from the simulation
region). The first term $L_m$ is the angular momentum flux carried
by matter, the second one $L_f$ is the angular momentum flux
carried by magnetic field.

The time dependence of the angular momentum fluxes carried by
matter $L_m$, by magnetic field $L_f$ and their sum $L=L_m+L_f$
are shown on Fig.~\ref{mu0507}a--c for different values of
$\zeta$.

For $\zeta=0$ (see Fig.~\ref{mu0507}a), the process of angular
momentum transport from the disk to the corona is quasi-periodic
with period equal approximately to ten rotation periods of the
disk inner part. After $t_*=45$ there is relaxation of the system
accompanied by small oscillations of the angular momentum fluxes.
During this time there no new plasmoids are generated. After
$t=65$, the reconnection process of the magnetic field lines is
resumed. It can be seen from the angular momentum fluxes
oscillations (see Fig.~\ref{mu0507}a).

The finite disk conductivity changes the course of events
(interaction between magnetic star and accretion disk). As it is
seen from Fig.~\ref{mu0507}a--c, the magnetic lines reconnection
lasts until some time $t_*$ (depending on $\zeta$). The more the
magnetic diffusivity $\zeta$, the later the plasmoids ejection
begins and the earlier ends. We should note that in case of an
imperfect disk conductivity the coronal magnetic field evolution
is qualitative similar to the perfect one. The reconnection of
magnetic field lines takes place at the time corresponding to the
maximum angular momentum outflow carried by magnetic field from
the disk. In contrast to a perfect disk conductivity ($\zeta=0$),
the activity of the reconnection slows down and the system relaxes
to such a state that no new plasmoids are generated. The disk
angular momentum carried by magnetic field gets balanced by
angular momentum carried by matter. Non-frozen in the disk
magnetic field lines move along the disk and it lead to the
increasing of the magnetic flux in the disk.

\section{Summary and conclusions}

The influence of the disk surface magnetic diffusivity on the
''star--corona--disk'' system evolution was studied. For all the
cases, poloidal field lines are pulled out and are reconnected
periodically approximately each tenth rotation period of the disk
inner bound. The magnetic field configuration has essentially
changed from initially dipole-like one.  The magnetic field lines
come out of its surface with large slope angle, fulfilling the
conditions for matter outflow from the disk, in the disk
differential rotation region.

The more the magnetic diffusivity, the more the distance from axis
and disk where the plasmoid ejection takes place. In the case of
imperfectly disk conductivity, magnetic field lines are no longer
frozen into the disk and, twisted by the star, they are shifted
outwards. The more surface magnetic diffusivity $\zeta$, the
slower the evolution takes place and more seldom plasmoids are
formed in corona.

In the case of an imperfectly disk conductivity, the coronal
magnetic field evolution is qualitative similar to the perfect
one. The reconnection of magnetic field lines takes place at the
time corresponding to the maximum angular momentum outflow
carried by magnetic field from the disk. In contrast to a perfect
disk conductivity, the activity of the reconnection slows down
and the system relaxes to such a state that no new plasmoids are
generated. The disk angular momentum carried by magnetic field
gets balanced by angular momentum carried by matter. Non-frozen in
the disk magnetic field lines move along the disk and it leads to
the increasing of the magnetic flux in the disk.

Summing up we can give following conclusions.
   \begin{enumerate}
      \item In the case of perfectly conducting disk the
evolution of the coronal magnetic field leads to quasi-periodic
outflow of the angular momentum from the disk. The interaction
process occurs with the reconnection of the magnetic field lines
and plasmoid ejections.
      \item In the case of imperfectly conducting disk the configuration of the magnetic field lines is
formed such that the angular momentum flux carried by the magnetic
field from the disk becomes balanced by the flux transported by
matter.
   \end{enumerate}

\begin{acknowledgements}
      Part of this work was supported by the
      \emph{Russian Foundation of Basic Research, RFBR\/} project
      number 06-02-16608, \emph{Scientific School Program} project number
      9399.2006.2 and \emph{RAS Presidium Program} number 14.
\end{acknowledgements}

\end{document}